\documentclass[a4paper]{jpconf}

\usepackage[]{graphicx}                                                                  

\newcommand{\ud}{\mathrm{d}}
 
\begin{document}
\title{Quantum multiresolution: tower of scales}

\author{Antonina N. Fedorova, Michael G. Zeitlin}

\address{IPME RAS, St.~Petersburg,
V.O. Bolshoj pr., 61, 199178, Russia\\
http://www.ipme.ru/zeitlin.html, http://mp.ipme.ru/zeitlin.html}

\ead{zeitlin@math.ipme.ru, anton@math.ipme.ru}

%%%%%%%%%%%%%%%%%%%%%%%%%%%%%%%%%%%%%%%%%%%%%%%%%%%%%%%%%%%%%%%%%%%%%%%%%%%%%%%%%
\begin{abstract}

We demonstrate the creation of nontrivial (meta) stable
states (patterns), localized, chaotic, entangled or decoherent,
from the basic localized modes in various collective models arising from the quantum hierarchy described
by Wigner-like equations.
The numerical simulation demonstrates the formation of various (meta) stable patterns or orbits generated
by internal hidden symmetry from generic high-localized fundamental modes.
In addition, we can control the type of behaviour on the pure algebraic level by means of properly
reduced algebraic systems (generalized dispersion relations).

\end{abstract}

\section{Introduction. New localized modes and patterns: why need we them?}

It is widely known that the currently available experimental techniques in the area of 
quantum physics as well 
as the present level of the understanding of phenomenological models, outstrips
the actual level of mathematical description.
Considering the problem of describing the really existing and/or realizable  states,
one should not expect that (gaussian) coherent states 
would be enough to characterize complex quantum phenomena. 
The complexity of a set of relevant states, including entangled (chaotic) ones 
is still far from being clearly understood and moreover from being realizable [1].
Our motivations arise from the following 
general questions [2]:
how can we represent a well localized and reasonable state in mathematically correct form?
is it possible to create entangled and other relevant states by means of these new localized 
building blocks?
The general idea is rather simple: it is well known that the
generating symmetry is the key ingredient of any modern
reasonable physical theory. 
Roughly speaking, the representation theory of the underlying 
(internal/hidden) symmetry (classical or quantum, finite 
or infinite dimensional, continuous 
or discrete) is the useful instrument for the description of (orbital) dynamics.
The proper representation theory is well known as ``local nonlinear harmonic analysis'',
in particular case of the simple underlying symmetry, affine group, aka wavelet analysis.
From our point of view the advantages of such approach are as follows:
{\bf i)} the natural realization of localized states in any proper functional realization of 
(Hilbert) space of states,
{\bf ii)} the hidden symmetry of a chosen realization of the functional model describes 
the (whole) spectrum of possible states via the so-called 
multiresolution decomposition.
Effects we are interested in are as follows: 
{\bf 1).} a hierarchy of internal/hidden 
scales (time, space, phase space);
{\bf 2).} non-perturbative multiscales: 
from slow to fast contributions,
from the coarser to the finer level of resolution/de\-composition;
{\bf 3)}.the coexistence of the levels of hierarchy of multiscale dynamics with transitions between scales;
{\bf 4).} the realization of the key features of the complex quantum 
world such as the existence of chaotic and/or entangled 
states with possible destruction in ``open/dissipative'' regimes due to interactions with
quantum/classical environment and transition to decoherent states.

N-particle Wig\-ner functions
allow to consider them as some quasiprobabilities. 
The full description for quantum ensemble can be done by the hierarchy
of functions (symbols):
$
W=\{W_s(x_1,\dots,x_s), s=0,1,2\dots\}
$
which are solutions of Wigner equations:
\begin{eqnarray}
\frac{\partial W_n}{\partial t}=-\frac{p}{m}\frac{\partial W_n}{\partial q}+
\sum^{\infty}_{\ell=0}\frac{(-1)^\ell(\hbar/2)^{2\ell}}{(2\ell+1)!}
\frac{\partial^{2\ell+1}U_n(q)}{\partial q^{2\ell+1}}
\frac{\partial^{2\ell+1}W_n}{\partial p^{2\ell+1}}.
\end{eqnarray}
The similar equations describe the important decoherence processes.

\section{Variational multiresolution representation}

We obtain our multiscale/multiresolution representations for solutions of Wig\-ner-like equations
(1) via the variational-wavelet approach [2] 
and represent the solutions as 
decomposition into localized eigenmodes  
related to the hidden underlying set of scales: 
$$
%\begin{eqnarray}
W_n(t,q,p)=\displaystyle\bigoplus^\infty_{i=i_c}W^i_n(t,q,p),
%\end{eqnarray}
$$
where value $i_c$ corresponds to the coarsest level of resolution
$c$ in 
the full multiresolution decomposition (MRA) [3]
of the underlying functional space:
$$
%\begin{equation}
V_c\subset V_{c+1}\subset V_{c+2}\subset\dots
%\end{equation}
$$
and $p=(p_1,p_2,\dots),\quad q=(q_1,q_2,\dots),\quad x_i=(p_1,q_1,\dots,p_i,q_i)$ 
are coordinates in phase space.
We introduce the Fock-like space structure on the whole space of internal hidden scales
$$
%\begin{eqnarray}
H=\bigoplus_i\bigotimes_n H^n_i
%\end{eqnarray}
$$
for the set of n-partial Wigner functions (states):
$$
%\begin{equation}
W^i=\{W^i_0,W^i_1(x_1;t),\dots,
W^i_N(x_1,\dots,x_N;t),\dots\},
%\end{equation}
$$
where
$W_p(x_1,\dots, x_p;t)\in H^p$,
$H^0=C,\quad H^p=L^2(R^{6p})$ (or any different proper functional spa\-ce), 
with the natural Fock space like norm: 
$$
%\begin{eqnarray}
(W,W)=W^2_0+
\sum_{i}\int W^2_i(x_1,\dots,x_i;t)\prod^i_{\ell=1}\mu_\ell.
%\end{eqnarray}
$$
First of all, we consider $W=W(t)$ as a function of time only,
$W\in L^2(R)$, via
multiresolution decomposition which naturally and efficiently introduces 
an infinite sequence of the underlying hidden scales.
We have the contribution to
the final result from each scale of resolution from the whole
infinite scale of spaces.
The closed subspace
$V_j (j\in {\bf Z})$ corresponds to  the level $j$ of resolution
and satisfies
the following properties:
let $D_j$ be the orthonormal complement of $V_j$ with respect to $V_{j+1}$: 
$
V_{j+1}=V_j\bigoplus D_j.
$
Then we have the following decomposition:
$$
%\begin{eqnarray}
\{W(t)\}=\bigoplus_{-\infty<j<\infty} D_j 
=\overline{V_c\displaystyle\bigoplus^\infty_{j=0} D_j},
%\end{eqnarray}
$$
in case when $V_c$ is the coarsest scale of resolution.
The subgroup of translations generates a basis for the fixed scale number:
$
{\rm span}_{k\in Z}\{2^{j/2}\Psi(2^jt-k)\}=D_j.
$
The whole basis is generated by the action of the full affine group [2], [3]:
$$
%\begin{eqnarray}
{\rm span}_{k\in Z, j\in Z}\{2^{j/2}\Psi(2^jt-k)\}=
{\rm span}_{k,j\in Z}\{\Psi_{j,k}\}
=\{W(t)\}.
%\end{eqnarray}
$$
After the construction of the multidimensional tensor product bases,  
the next key point is 
the so-called Fast Wavelet Transform (FWT) [3], 
demonstrating that for a large class of
operators the wavelet functions are a good 
approximation for true eigenvectors; and the corresponding 
matrices are almost diagonal. 
We have the simple linear para\-met\-rization of the
matrix representation of  our operators in the localized wavelet bases
and of the action of
these operators on arbitrary vectors/states in the proper functional space.
FWT provides  the maximum sparse and useful form for the wide classes 
of operators.
After that,
we can obtain our multiscale\-/mul\-ti\-re\-so\-lu\-ti\-on 
representations for observables (symbols), states, partitions
via the variational approaches. 
Let $L$ be an arbitrary (non)li\-ne\-ar dif\-fe\-ren\-ti\-al\-/\-in\-teg\-ral operator 
 with matrix dimension $d$
(finite or infinite), 
which acts on some set of functions
from $L^2(\Omega^{\otimes^n})$:  
$\quad\Psi\equiv\Psi(t,x_1,x_2,\dots)=\Big(\Psi^1(t,x_1,x_2,\dots), \dots$,
$\Psi^d(t,x_1,x_2,\dots)\Big)$,
 $\quad x_i\in\Omega\subset{\bf R}^6$, $n$ is a number of particles:
\begin{eqnarray*}
L\Psi&\equiv& L(Q,t,x_i)\Psi(t,x_i)=0,\\
Q&\equiv& Q_{d_0,d_1,d_2,\dots}(t,x_1,x_2,\dots,
\partial /\partial t,\partial /\partial x_1,
\partial /\partial x_2,\dots,
\int \mu_k)
\\
&=&\sum_{i_0,i_1,i_2,\dots=1}^{d_0,d_1,d_2,\dots}
q_{i_0i_1i_2\dots}(t,x_1,x_2,\dots)
\Big(\frac{\partial}{\partial t}\Big)^{i_0}\Big(\frac{\partial}{\partial x_1}\Big)^{i_1}
\Big(\frac{\partial}{\partial x_2}\Big)^{i_2}\dots\int\mu_k.
\end{eqnarray*}
Let us consider the $N$ mode approximation:
$$
%\begin{eqnarray}
\Psi^N(t,x_1,x_2,\dots)=
\sum^N_{i_0,i_1,i_2,\dots=1}a_{i_0i_1i_2\dots}
 A_{i_0}\otimes 
B_{i_1}\otimes C_{i_2}\dots(t,x_1,x_2,\dots)
%\end{eqnarray}
.
$$
We will determine the expansion coefficients from the following conditions,
Generalized Dispersion Relation,
related to the proper choosing of variational approach:
\begin{eqnarray}
&&\ell^N_{k_0,k_1,k_2,\dots}\equiv 
\int(L\Psi^N)A_{k_0}(t)B_{k_1}(x_1)C_{k_2}(x_2)\ud t\ud x_1\ud x_2\dots=0.
\end{eqnarray}
Thus, we have exactly $dN^n$ algebraical equations for  $dN^n$ unknowns 
$a_{i_0,i_1,\dots}$.
This variational ap\-proach reduces the initial problem 
to the problem of solution 
of functional equations at the first stage and 
some algebraical problems at the second one.
It allows to unify the multiresolution expansion with variational 
construction. 
As a result, the solution is parametrized by the solutions of two sets of 
reduced algebraical
problems, one is linear or nonlinear
(depending on the structure of the generic operator $L$) and the rest are linear
problems related to the computation of the coefficients of reduced 
algebraic equations. It is also related to the choice of exact measure of localization
(including the class of smoothness), which is proper for our set-up.
These coefficients can be found  via functional/algebraic methods
by using the
compactly supported wavelet basis or any other wavelet families.
As a result, the solution of the hierarchies as in c-
as in q-region, has the 
following mul\-ti\-sca\-le or mul\-ti\-re\-so\-lu\-ti\-on decomposition via 
nonlinear lo\-ca\-li\-zed eigenmodes 
{\setlength\arraycolsep{0pt}
\begin{eqnarray}
&&W(t,x_1,x_2,\dots)=
\sum_{(i,j)\in Z^2}a_{ij}U^i\otimes V^j(t,x_1,\dots),\nonumber\\
&&V^j(t)=
V_N^{j,slow}(t)+\sum_{l\geq N}V^j_l(\omega_lt), \ \omega_l\sim 2^l, \nonumber\\
&&U^i(x_s)=
U_M^{i,slow}(x_s)+\sum_{m\geq M}U^i_m(k^{s}_mx_s), \ k^{s}_m\sim 2^m,
\end{eqnarray}
which corresponds to the full multiresolution expansion in all underlying time/space 
scales.
The formulas (3) give the expansion into a slow part
and fast oscillating parts for arbitrary $N, M$.  So, we may move
from the coarse scales of resolution to the 
finest ones for obtaining more detailed information about the dynamical process.
In this way, one obtains contributions to the full solution
from each scale of resolution or each time/space scale or from each nonlinear eigenmode. 
Formulas (3) do not use perturbation
techniques or linearization procedures.
Numerical calculations are based on compactly supported
wavelets and wavelet packets and on the evaluation of  
accuracy on 
the level $N$ of the corresponding cut-off of the full system 
regarding Fock-like norm:
$
\|W^{N+1}-W^{N}\|\leq\varepsilon.
$

\section{Conclusions}

By using proper high-localized bases on orbits generated by actions of internal hidden symmetries of 
underlying functional spaces, we can describe and classify the full zoo of patterns with 
non-trivial behaviour including localized (coherent) structures in      
quantum systems with complicated behaviour (Figs.~1, 2).
The numerical simulation demonstrates the formation of various (meta) stable patterns or orbits 
generated by internal hidden symmetry from generic 
high-localized fundamental modes.
These (nonlinear) eigenmodes are more realistic for the modeling of 
classical/quantum dynamical process  than the (linear) gaussian-like
coherent states. 
Here we mention only the best convergence properties of the expansions 
based on wavelet packets, which  realize the minimal Shannon entropy property
and the exponential control of the convergence of expansions like (3).
Figs.~1, 2 demonstrate the steps of (hidden) multiscale resolution, starting from coarse--graining, 
during the full quantum interaction/evolution  
of entangled states leading to the growth of the degree of complexity (entanglement) of the quantum state.
It should be noted that
we can control the type of behaviour on the level of the reduced algebraic 
system (Generalized Dispersion Relation) (2) [2]. 
\begin{figure}[h]
\begin{minipage}{18pc}
\includegraphics[width=18pc]{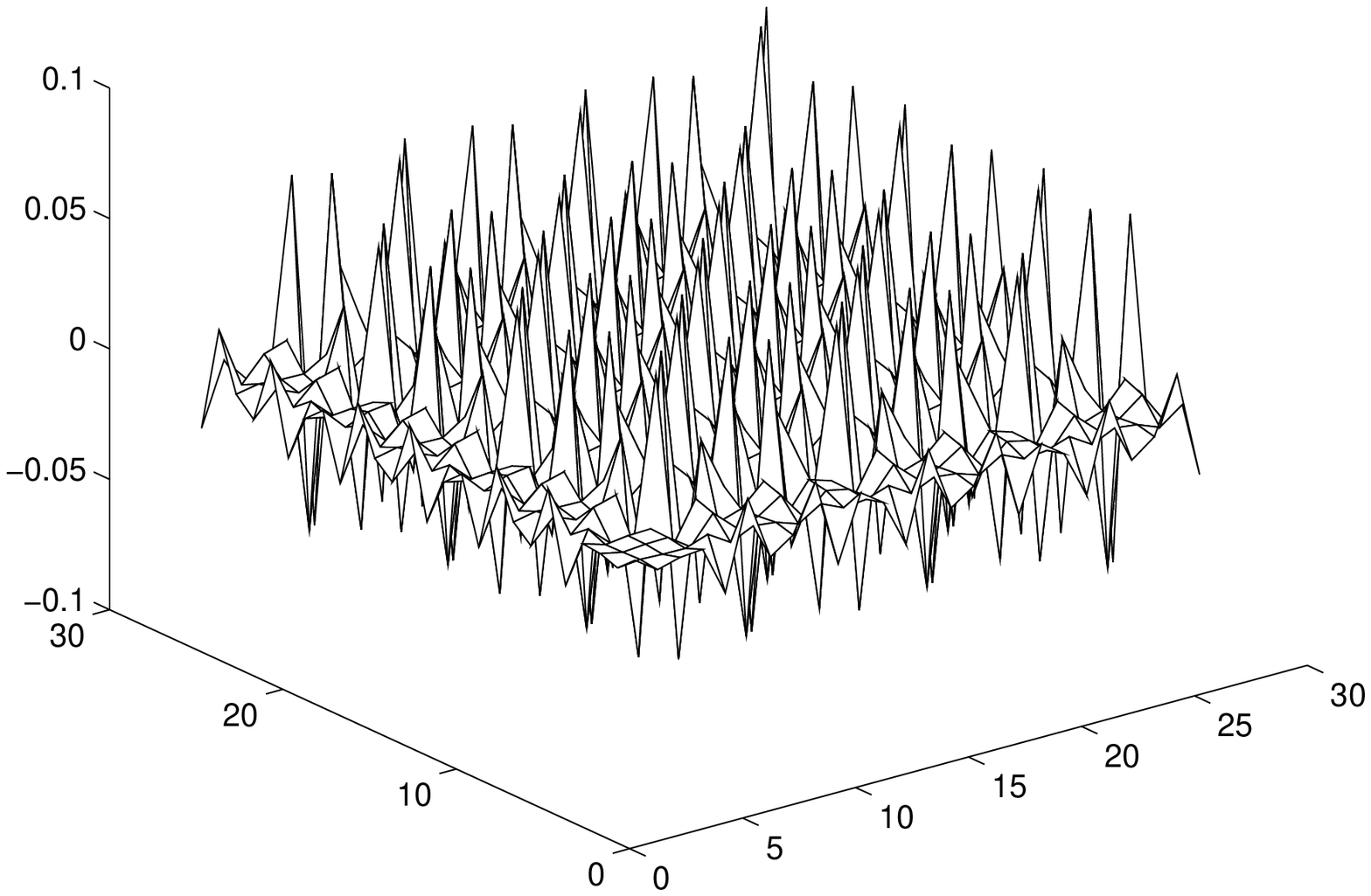}
\caption{Entangled Wigner function.}
\end{minipage}\hspace{2pc}%
\begin{minipage}{18pc}
\includegraphics[width=18pc]{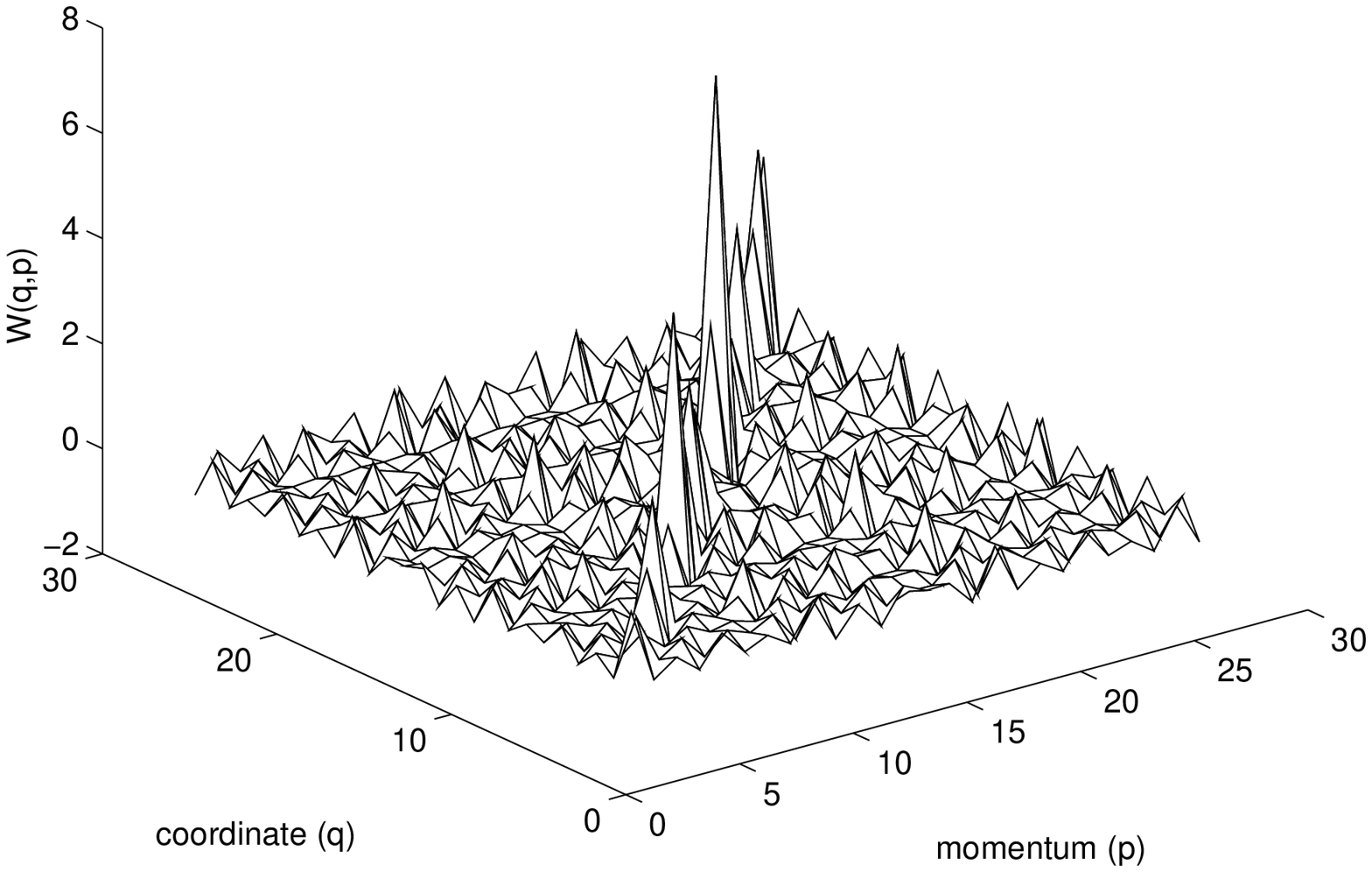}
\caption{Localized (decoherent) pattern: (waveleton) Wigner function.}
\end{minipage}
\end{figure}

\end{document}